\documentstyle[12pt]{article}

\def\be{\begin{equation}}
\def\ee{\end{equation}}
\def\bea{\begin{eqnarray}}
\def\eea{\end{eqnarray}}

\def\beq{\begin{equation}}   \def\eeq{\end{equation}}
\def\bea{\begin{eqnarray}}   \def\eea{\end{eqnarray}}

\newcommand{\matel}[3]{\langle #1|#2|#3\rangle}

\begin{document}
\begin{flushright}
UND-HEP-01-BIG\hspace*{.2em}02\\
\end{flushright}


\vspace{.3cm}
\begin{center} \Large
{\bf
{CHARM PHYSICS AND THE POOR SLEEPER'S IMPATIENCE}}
\footnote{Invited talk given at the 4th International
Conference on B Physics and CP Violation, Ise, Japan,
Feb. 19 - 23, 2001}
\\
\end{center}
\vspace*{.3cm}
\begin{center} {\Large
I. I. Bigi }\\
\vspace{.4cm}
{\normalsize
{\it Physics Dept.,
Univ. of Notre Dame du
Lac, Notre Dame, IN 46556, U.S.A.} }
\\
\vspace{.3cm}
{\it e-mail address: bigi@undhep.hep.nd.edu }
\vspace*{0.4cm}

{\Large{\bf Abstract}}
\end{center}
After a short review of
the theoretical tools available to describe heavy flavour
physics I sketch the present profile of the weak dynamics of charm
hadrons with respect to lifetimes, oscillations and CP violation.
I argue that comprehensive studies of charm decays provide novel
portals to New Physics and suggest some benchmark figures for
desirable sensitivities.


\tableofcontents
\section{Introduction}
\label{INTRO}

There is a widespread perception that while charm physics had its days,
those days are in the past, it has little to offer in new
insights and all that remains is polishing the data. This is, however,
a short-sighted view with respect to both production and decay of charm
hadrons. I will focus on the latter to emphasize the following points:
\begin{itemize}
\item
They provide us with a test bed for QCD technologies.
\item
Charm transitions are a unique portal for obtaining a novel access to the
flavour problem with the experimental situation being a priori
favourable.

\end{itemize}
Accordingly I will address the following topics: first I will give a
lightning
update on the status of QCD technologies relevant for heavy flavour
decays; after sketching charm's present profile I will stress its promise
of revealing New Physics and combine it with  an appeal to embark onto a
comprehensive New Phenomenology emphasizing
$D^0 - \bar D^0$ oscillations and CP violation.

\section{QCD Technologies}
\label{QCD}

While we have no solution of full QCD, we do have theoretical
technologies inferred from QCD that allow us to deal with
nonperturbative dynamics in  special situations. Those are chiral
perturbation theory for pion and kaon dynamics and heavy quark
expansions (HQE). The latter apply to some aspects of
the dynamics of beauty hadrons and possibly of charm hadrons as
well. However
since the charm quark mass exceeds ordinary hadronic scales
by  a moderate margin only, one can expect at best a
semi-quantitative description there.

Simulating QCD on the lattice represents a technology of wide reach.
In principle lattice QCD could work its way up to the charm scale from
below. However the considerable advances achieved recently on the
lattice with respect to heavy flavour physics were not based on such a
`brute-force' approach, but on a judicious choice of $1/m_Q$
expansions \cite{HASHIMOTO,AOKI}.
The feedback between HQE and lattice QCD will yield even more gains
in the future.

\subsection{Heavy Quark Expansions}

In HQE one describes an observable $\gamma$ for a hadron $H_Q$
-- be it a total rate or a distribution --
through an expansion in inverse powers of the heavy
{\em quark} mass \cite{HQT}
$m_Q$:
\beq
\gamma(E) = \sum _i c_i(\alpha _S, E) (\Lambda _i/m_Q)^i \; ;
\eeq
$E$ denotes the relevant energy scale.

The crucial question in this context is whether an observable
can be described through an operator product expansion (OPE) or not.
Essential tools are provided by sum rules \cite{OPTICAL}
\beq
\int _0^{\infty} dEw(E) \gamma (E)|_{hadrons} =
\int _0^{\infty} dEw(E) \gamma (E)|_{quarks}
\eeq
stating that the integral of such observable $\gamma$ weighted
by some function $w(E)$ has to be equal when expressed in terms of
hadronic or quark degrees of freedom.

These methods are applied to {\em in}clusive transitions -- lifetimes,
semileptonic branching ratios, lepton spectra etc. -- and
{\em ex}clusive observables like semileptonic form factors and branching
ratios for nonleptonic two-body modes. I would like to add here that
there are several reasons why the recently suggested methods for
$B\to M_1M_2$ \cite{PQCD} are hard to justify for charm decays.
Nevertheless one  should try them there anyway!

When calculating a rate on the quark-gluon level
{\em quark-hadron duality} (or duality for short) is invoked to equate
the result with what one should get for the corresponding process
expressed in hadronic quantities.
Such duality represents a very natural concept. For the hadronic final
state forms in two steps: a {\em hard} process controled by a time
scale $1/m_Q$ is followed by {\em soft} hadronization characterized by
a time scale $ \Lambda$ in the rest frame of the heavy quark which
gets time dilated into $\sim m_Q/ \Lambda ^2$ in the, say, c.m.
frame. Since $m_Q/ \Lambda ^2 \gg 1/m_Q$ the gross features of the
process -- total rates, energy flows etc. -- are determined by the
first step. Duality thus has to be exact at asymptotic scales; yet
at finite scales there are corrections.

These features can nicely be illustrated by a quantum mechanical model
involving a potential $V(\vec x)$: while the local properties of the
potential determine integrated rates
\footnote{This connection can be broken, if the potential contains
singularities at finite distances. This feature can be reproduced in a
quantum field theory through instanton effects \cite{MISHA}.}
,
the asymptotic features (like  confining or not) control the
specifics of the final state.

While we have no complete theory yet for the limitations to duality, we
have moved beyond a merely folkloric stage.
\begin{itemize}
\item
We know that the duality violations at finite scales
depend on the process under study.
\item
We have identified the mathematical portals for duality violations:
the OPE constructed in the Euclidean regime cannot reproduce terms
like exp$(-m_Q/\Lambda)$ that are exponentially suppressed there.
However upon analytic continuation into the Minskowski domain they
get transmogrified into terms like sin$(m_Q/ \Lambda)$, which by
themselves are not suppressed. It turns out though that duality violating
terms are power suppressed, i.e. of the form
sin$(m_Q/\bar \Lambda)/m_K^{k_i}$ with a positive power $k_i$ that
depends on the reaction.
\item
The fundamental question is whether one can base the description on
an OPE or not rather than whether one deals with nonleptonic versus
semileptonic transitions or whether one considers local duality.
One expects duality violations to be numerically larger in the former
than the latter class of processes, but not as a matter of principle!

\item
One particular and obvious problem for the charm sector: the expansion
parameter $\Lambda _i/m_c$ is not much smaller than unity since
$m_c(m_c) = 1.25 \pm 0.1$ GeV. {\em At best} this introduces sizeable
numerical uncertainties; {\em at worst} it could signal the breakdown
of duality at or near the charm scale.

\end{itemize}
These insights are based on two types of sources:
\begin{itemize}
\item
We have developed a good understanding of the physical origins of
duality violations as due to hadronic thresholds, the presence of
`distant' cuts and the $1/m_c$ expansions \cite{URALTSEVJOFFE}.
\item
Very extensive and detailed studies in model field theories like the
't Hooft model have been performed over the last few years
\cite{THOOFT}.

\end{itemize}
The final arbiter will be provided by data, of course, namely by
{\em over}constraints in measurements. One topical example is the
beauty quark mass which can be extracted from the $\Upsilon (4S)$
mass and from the shape of the lepton spectra in semileptonic
$B$ decays. Present results are quite encouraging in that respect
\cite{OSAKA}.

\subsection{Lattice QCD}

Lattice QCD, which was originally introduced to prove confinement
and bring hadronic spectroscopy under theoretical control, is now
making major contributions to heavy flavour physics -- with partially
unquenched results. For the $D_s$ decay constant one finds
for two active flavours \cite{AOKI}
\beq
f(D_s) =
255 \pm 30 \; {\rm MeV} \; \;  {\rm lattice\, QCD}
\label{FDS}
\eeq
to be compared with what one infers from a world average of data
on $ D_s \to \mu \nu$ \cite{MARINELLI}:
$\langle  f(D_s)\rangle = 269 \pm 22 \; {\rm MeV}$;
the experimental error is probably on the optimistic side.

In the future one expects to measure the $D_s^+$ and $D^+$ decay
constants with 5 - 10 \% accuracy at the beauty factories; a
$\tau$-charm factory would allow a 2-3 \% measurement
\cite{ALEXANDER}, which could be
fully utilized since absolute branching ratios can be determined
with the necessary accuracy at the same time. On the theoretical side
a lattice study with full unquenching that treats charm quarks as
dynamic rather than static entities is not utopian; likewise for the
form factors in exclusive semileptonic charm decays. Charm decays
will thus provide a rich lab for quantitative tests of lattice QCD.

More generally, the synergies produced by the feedback between HQE
and lattice QCD and the calibration provided by the charm data
will yield important lessons on QCD probing and hopefully extending
our theoretical control over nonperturbative dynamics. Beyond being
a very worthwhile motivation in itself, it will strengthen searches
for New Physics.

\section{Present Profile of the Weak Dynamics of Charm}

\subsection{Lifetimes}

As explained at BCP3 \cite{BIGIBCP3} HQE produce a remarkably successful
description of both the pattern and the numbers in the lifetime ratios.
The HQE provide a rationale for most of the  phenomenological concepts
introduced before -- like PI, WA etc. --  as effects of order $1/m_Q^3$.
Since it represents a self-consistent  framework, it is more definitive
about those concepts. For example WA  has to be a nonleading effect
although it could still be significant.

HQE also explain -- as a new and highly welcome
feature -- the absolute sizes of the semileptonic branching ratios as
effects of order $1/m_Q^2$ -- something that previous models could not.

The one new result of the last two years is \cite{GOLUTVIN,TANAKA}
\beq
\tau (D_s)/\tau (D^0) = 1.18 \pm 0.02
\eeq
rather than the previous world average of $1.125 \pm 0.042$. It confirms
that WA is {\em not the leading} source of lifetimes differences
among charm mesons; at the same time it shows WA to be still
significant at the about 20 \% level, as expected. It leads to the
further question whether this difference in total widths can be traced
to certain classes of exclusive channels.

The accuracy in the data on $\tau (\Xi _c^{0,+})$ and
$\tau (\Omega _c)$ leaves much to be desired before
firm conclusions can be based on them. One should also remember that
the semileptonic {\em widths} of charm baryons do not mirror their
lifetime ratios since the semileptonic widths are {\em not}
universal for baryons \cite{VOLSL}.

\subsection{Cabibbo Hierarchy}

The full range of the Cabibbo pattern has been observed in
nonleptronic as well as semileptonic transitions.

Imposing three-family unitarity leads to numerically very precise
values of the CKM parameters:
\beq
|V(cs)| = 0.9742 \pm 0.0008 \; , \; |V(cd)| = 0.222 \pm 0.003
\eeq
Without that constraint the values are less
precise:
\beq
|V(cs)| = 0.880 \pm 0.096 \; , \; |V(cd)| = 0.226 \pm 0.007
\eeq
As far as $|V(cs)|$ is concerned the main information
from semileptonic $D$ decays is augmented by findings from charm
production in deep inelastic neutrino scattering; for
$|V(cd)|$ it is the other way around. A recent
OPAL analysis of $W\to$ charm jets obtains
\beq
|V(cs)| = 0.969 \pm 0.058
\eeq

\subsection{Data on $D^0 - \bar D^0$ Oscillations}

Oscillations are described by the normalized mass and width
differences:
$x_D \equiv \frac{\Delta M_D}{\Gamma _D}$,
$y_D \equiv \frac{\Delta \Gamma}{2\Gamma _D}$.
The experimental landscape is described by the following numbers
\cite{CLEOCP,TANAKA}:
\bea
x_D &\leq& 0.03   \\
 y_D &=&
\left\{
\begin{array}{l}
(0.8 \pm 2.9 \pm 1.0) \% \; \; {\rm E791} \\
(3.42 \pm 1.39 \pm 0.74) \% \; \; {\rm FOCUS} \\
(1.16^{+1.67}_{-1.65})\% \; \; {\rm BELLE} \\
(- 1.1 \pm 2.5 \pm 1.4) \% \; \; {\rm CLEO}
\label{DATAOSC}
\end{array}
\right.
\\
y_D^{\prime} &=& (-2.5 ^{+1.4}_{-1.6} \pm 0.3)\% \; \;
{\rm CLEO}
\; .
\eea
$y_D^{\prime}$ is extracted from fitting
a general
lifetime evolution to
$D^0(t) \to K^+\pi ^-$ and depends on the
strong rescattering phase $\delta$ between $D^0 \to K^-\pi^+$ and
$D^0 \to K^+\pi^-$:
$y_D^{\prime}
= -x_D{\rm sin}\delta + y_D {\rm cos}\delta $. It
could differ substantially
from
$y_D$ if that phase were
sufficiently large \cite{PETROV}.
All measurements are still consistent with zero.

\subsection{CP Asymmetries -- Data}

Data are summarized in Table \ref{CPAS} \cite{PEDRINI,CLEOCP}.
\begin{table}
\begin{tabular} {|l|l|l|}
\hline
 channel& $D^0 \to K^+ K^-$ & $D^0 \to \pi^+ \pi^-$ \\
\hline
\hline
E 791 & $- 1.0 \pm 4.9 \pm 1.2$\% & $- 4.9 \pm 7.8 \pm 3.0$\%
 \\
\hline
CLEO & $ 0.05 \pm 2.18 \pm 0.84$\% &
$ 1.95 \pm 3.22 \pm 0.84$\%   \\
\hline
FOCUS & $- 0.1 \pm 2.2 \pm 1.5 $\%  &
$ 4.8 \pm 3.9 \pm 2.5 $\%
 \\
\hline
\hline
\hline
channel & $D^{\pm} \to K^+ K^-\pi ^{\pm}$  &\\
\hline
\hline
E 791 & $- 1.4 \pm 2.9$\% &
 \\
\hline
FOCUS &  $ 0.6 \pm 1.1 \pm 0.5 $\%  &
 \\
\hline
\end{tabular}
\centering
\caption{Data on direct CP asymmetries in $D$ decays}
\label{CPAS}
\end{table}
The experimental sensitivity has increased significantly
to put us within striking distance of the 1\% level. Yet
the numbers are still consistent with zero.

\section{Charm Decays -- Novel Portals to New Physics}

Obviously there is unfinished business in charm physics: one wants to
measure {\em absolute} branching ratios more precisely, in particular for
$D_s$ and charm baryons; likewise for $\Xi _c^{0,+}$ and $\Omega _c$
lifetimes and semileptonic branching ratios. As already mentioned, more
accurate data on $D^+,\, D_s^+ \to l^+\nu$ are very desirable -- as are
post-MARK III data on lepton {\em spectra} in {\em inclusive}
semileptonic charm decays. All these can provide important inputs to
beauty studies and some can give us important lessons on QCD as well.

But that is not the end of it! There is a wide-spread conviction in the
community that the SM is incomplete, and our efforts are focussed
on uncovering New Physics. It seems to me that charm decays have
a good potential to reveal manifestations of New Physics that
might not be manifest in beauty decays. For charm quarks are the
only up-type quark allowing a full range of indirect searches
for New Physics. While $D^0 -\bar D^0$ oscillations are slow,
$T^0 - \bar T^0$ oscillations
cannot occur at all, nor can CP violation there, since
top
quarks decay before they can hadronize \cite{DOK}.
Direct CP violation can emerge in exclusive modes that command
decent branching ratios for charm, but are really tiny for
top with little coherence left.

Finally charm decays proceed in an environment populated with
many resonances which induce final state interactions (FSI) of
great vibrancy. While this feature complicates the interpretations
of a signal (or lack thereof) in terms of microscopic quantities, it
is optimal for getting an observable signal. In that sense it
should be viewed as a glass half full rather than half empty.

Charm hadrons provide several practical advantages and opportunities:
their production rates are relatively large; they possess long
lifetimes and
$D^* \to D\pi$ decays provide as good a flavour tag as one can have.

This leads to my basic contention: charm transitions are a
{\em unique} portal for obtaining a {\em novel} access to the
{\em flavour} problem with the
{\em experimental situation being a priori mostly favourable}!

\subsection{$D^0 - \bar D^0$ Oscillations -- Revisited}

While all present data are consistent with
both $x_D$ and $y_D$ being zero, we have to examine how significant
that statement is, i.e. what the SM expectations are.

With $D^0 \to f \to \bar D^0$ transition amplitudes being proportional
to sin$\theta _C^2$ one has $x_D, y_D \leq 0.05$; furthermore
in the limit of $SU(3)_{Fl}$ symmetry those amplitudes have to vanish.
However a priori one cannot count on that being a very strong suppression
for the real world; thus
$ x_D, \; y_D \sim {\cal O}(0.01)$ represents a
conservative SM bound. On general grounds I find it unlikely
-- though mathematically possible -- that New Physics could
overcome the Cabibbo bound significantly. Comparing this general bound
on the oscillation variables to the data listed
in Eq.(\ref{DATAOSC}), I conclude the hunt for New Physics realistically
has only just begun!

One can give a more sophisticated SM estimate for
$x_D$, $y_D$. There exists an extensive literature on it
\cite{BURDMAN}; however
some relevant features were missed for a long time. Quark box diagrams
yield tiny contributions only:
\beq
x_D({\rm box}) \sim {\rm few} \; \times 10^{-5}
\label{BOX}
\eeq
Various schemes are then invoked  to describe selected hadronic
intermediate states to guestimate the impact of long distance
dynamics:
\beq
x_D(LD), \; y_D(LD) \sim 10^{-4} - 10^{-3}
\label{LD}
\eeq
Recently a new analysis
\cite{DOSC} has been given based on an OPE providing a
systematic treatment in powers of $1/m_c$, the GIM factors $m_s$
and the CKM parameters. It finds that the the GIM suppression
by a factor of $(m_s/m_c)^4$, which is behind the result stated in
Eq.(\ref{BOX}) is {\em untypically severe} \cite{GEORGI}. It was found
that there are contributions with gentle GIM factors
proportional to $m_s^2/\mu _{had}^2$ or even $m_s/\mu _{had}$.
They are due to higher-dimensional operators and thus accompanied
by higher powers of $1/m_c$. Since those are not greatly suppressed,
contributions of formally higher order in $1/m_c$ can become numerically
leading if they are of lower order in $m_s$. These contributions
are actually due to condensate terms in the OPE, namely
$\matel{0}{\bar qq}{0}$ etc. On the {\em conceptual} side we have
achieved significant progress: it is again the OPE that allows
to incorporate nonperturbative dynamics from the start in a
self-consistent way. {\em Numerically} there is no decisive change,
although the numbers are somewhat larger:
\beq
x_D(SM)|_{OPE}, \; y_D(SM)|_{OPE} \; \sim \; {\cal O}(10^{-3})
\eeq
However one realizes that it is rather unlikely that the uncertainties
can significantly be reduced, since the values depend on a high
power of some hadronic quantities.

The crucial question is: does duality hold at the charm scale
for $x_D$ and $y_D$? Those two observables are sensitive to different
aspects of $\Delta C=2$ dynamics: (i) The normalized width difference is
determined by on-shell transitions and has very little chance to be
affected by New Physics; on the other hand it can be strongly affected
by a near-by resonance. Whether duality has any validity for the
observable $y_D$ is quite unclear a priori.
(ii) The mass difference on the
other hand is controlled by virtual transitions. Thus it has
a good chance to be shaped by New Physics leading to
$x_D \sim {\cal O}({\rm few}\%)$; at the same time
it involves more smearing than $y_D$; therefore duality has a much
better chance to apply approximately to $x_D$ than to $y_D$.

If data revealed $y_D \ll x_D \sim 1\%$ we would have a strong case to
infer the intervention of New Physics. If on the other hand
$y_D \sim 1\%$ -- as hinted at by the FOCUS data -- then two scenarios
could arise:
if $x_D \leq {\rm few}\times 10^{-3}$ were found, one would infer
that the $1/m_c$ expansion within the SM yields a correct
semiquantitative result while blaming the "large" value for
$y_D$ on a sizeable and not totally surprising violation of
duality. If, however, $x_D \sim 0.01$ would emerge, we would face a
theoretical conundrum: an interpretation ascribing this to
New Physics would hardly be convincing since $x_D \sim y_D$.
To base a case for New Physics solely on the observation of
$D^0 - \bar D^0$ oscillations is thus of uncertain value.

\subsection{CP Violation -- Expectations}

\noindent (i) Direct CP Violation in Partial Widths

For an asymmetry to become observable
between CP conjugate partial widths, one needs two coherent
amplitudes with a relative {\em weak} phase and a nontrivial
strong phase shift.

In Cabibbo favoured as well as in doubly Cabibbo suppressed
channels those requirements can be met with New Physics only. There is
one exception to this general statement  \cite{YAMA}: the transition
$D^{\pm} \to K_S \pi ^{\pm}$ reflects the interference between
$D^{+} \to \bar K^0 \pi ^+$ and $D^+ \to K^0 \pi ^+$ which
are Cabibbo favoured and doubly Cabibbo suppressed, respectively.
Furthermore in all likelihood those two amplitudes will exhibit
different phase shifts since they differ in their isospin
content.
The known CP impurity in the $K_S$ state induces a
difference without any theory uncertainty:
$$
\frac{\Gamma (D^+ \to K_S \pi ^+) - \Gamma (D^- \to K_S \pi ^-)}
{\Gamma (D^+ \to K_S \pi ^+) + \Gamma (D^- \to K_S \pi ^-)} =
-2{\rm Re}\epsilon _K
$$
\beq
\simeq - 3.3 \cdot 10^{-3}
\label{DKSSM}
\eeq
In that case the same asymmetry both in magnitude as well
as sign arises for the experimentally much more challenging
final state with a $K_L$.
If on the other hand New Physics is present in $\Delta C=1$ dynamics,
most likely in the doubly Cabibbo transition, then both the
sign and the
size of an asymmetry can be different from the number in Eq.(\ref{DKSSM}),
and by itself
it
would make a contribution of the opposite sign to the asymmetry in
$D^+ \to K_L\pi ^+$ vs. $D^- \to K_L\pi ^-$.

Searching for {\em direct} CP violation in
Cabibbo suppressed $D$ decays as a sign for New Physics would also
represent a very complex challenge: within the KM description one expects
to find some asymmetries of order 0.1 \%; yet it would be hard
to conclusively rule out some more or less accidental enhancement due to a
resonance etc. raising an asymmetry to the 1\% level.
Observing a CP
asymmetry in charm decays would certainly be a first rate discovery even
irrespective of its theoretical interpretation.
Yet to make a case that a
signal in a singly  Cabibbo suppressed mode reveals New Physics is quite
iffy. In all  likelihood one has to analyze at least several channels
with comparable  sensitivity to acquire a measure of confidence in one's
interpretation.

\noindent (ii) Direct CP Violation in Final State Distributions

For channels with two pseudoscalar mesons or a pseudoscalar and a vector
meson a CP asymmetry can manifest itself only in a difference between
the two partial widths. If, however, the final state
is more complex -- being made up by three pseudoscalar or two
vector mesons etc. -- then it contains more dynamical information than
expressed by its partial width, and CP violation can emerge also through
asymmetries in final state distributions. One general comment
still applies: since also such CP asymmetries require the
interference of two weak amplitudes, within the SM
they can occur in Cabibbo
suppressed modes only.

In the simplest such scenario one compares CP conjugate
{\em Dalitz plots}. It is quite
possible that different regions of a Dalitz plot exhibit CP
asymmetries of varying signs that largely cancel each other when
one integrates over the whole phase space. I.e., subdomains of the
Dalitz plot could contain considerably larger CP asymmetries
than the integrated partial width.

Once a Dalitz plot is fully understood with all its resonance and
non-resonance contributions including their strong phases, one has a
powerful and sensitive new probe. This is not an easy goal to
achieve, though, in particular when looking for effects that
presumably are not large. It might be more promising
as a practical matter to start out with a more euristic approach.
I.e., in the spirit of Yogi Berra one can start a search for
CP asymmetries by just looking at conjugate Dalitz plots. One simple
strategy would be to focus on an area  with a resonance band and analyze
the density in stripes {\em across} the  resonance as to whether there is
a difference in CP conjugate plots.

For more complex final states containing
four pseudoscalar mesons etc. other probes have to be
employed.  Consider for example
$
D^0 \to K^+K^- \pi ^+ \pi ^- \; ,
$
where one can form a T-odd correlation with the momenta:
$
C_T \equiv \langle \vec p_{K^+}\cdot
(\vec p_{\pi^+}\times \vec p_{\pi^-})\rangle
$.
Under time reversal T one has
$
C_T \to - C_T
$
hence the name `T-odd'. Yet $C_T \neq 0$ does not necessarily
establish T violation. Since time reversal is implemented
by an {\em anti}unitary operator, $C_T \neq 0$ can be induced by
final state interactions (FSI). While in contrast to the situation
with partial width differences FSI are not required to produce
an effect, they can act as an `imposter' here, i.e. induce a T-odd
correlation with T-invariant dynamics. This ambiguity can unequivoally
be resolved by measuring
$
\bar C_T \equiv \langle \vec p_{K^-}\cdot
(\vec p_{\pi^-}\times \vec p_{\pi^+})\rangle
$
in $\bar D^0 \to K^+K^- \pi ^+ \pi ^- $; finding
$
C_T \neq - \bar C_T
$
establishes CP violation without further ado.

Decays of {\em polarized} charm baryons provide us with a
similar class of observables; e.g., in
$\Lambda _c \Uparrow \; \to p \pi ^+\pi ^-$, one can analyse the
T-odd correlation $\langle \vec \sigma _{\Lambda _c}
\cdot (\vec p_{\pi ^+} \times \vec p_{\pi ^-})\rangle$ \cite{BENSON}.

\noindent (iii) CP violation involving $D^0 - \bar D^0$ oscillations

The interpretation is much clearer once one finds a CP
asymmetry that involves oscillations; i.e., one compares
the time evolution of transitions like $D^0(t) \to K_S \phi$,
$K^+ K^-$, $\pi ^+ \pi ^-$ and/or
$D^0(t) \to K^+ \pi ^-$ with their CP conjugate channels. A
difference for a final state $f$ would depend
on the product
\beq
{\rm sin}(\Delta m_D t) \cdot {\rm Im} \frac{q}{p}
[T(\bar D\to f)/T(D\to \bar f)] \; .
\eeq
With  both factors being
$\sim
{\cal O}(10^{-3})$ in the SM with the KM ansatz
one predicts a practically zero
asymmetry $\leq 10^{-5}$. Yet
New Physics could quite conceivably generate
considerably larger values, namely
$x_D \sim {\cal O}(0.01)$,
Im$\frac{q}{p}
[T(\bar D\to f)/T(D\to \bar f)] \sim  {\cal O}(0.1)$
leading to an asymmetry of ${\cal O}(10^{-3})$.
One should note that the
oscillation dependant term is linear in the
small quantity $x_D$
\beq
{\rm sin}\Delta m_D t
\simeq x_D t /\tau _D
\eeq
in contrast to $r_D$ which is
quadratic:
\beq
r_D \equiv \frac{D^0 \to l^-X}{D^0 \to l^+X}
\simeq \frac{x_D^2 + y_D^2}{2}
\eeq
It would be very hard to see $r_D = 10^{-4}$ in CP insensitive
rates. It could then well happen that $D^0 - \bar D^0$
oscillations are first discovered in such CP asymmetries!

\section{Summary and Outlook}

We have learnt many important lessons from charm studies. Yet even so,
they do not represent a closed chapter.

On one hand charm physics
can teach us many more important lessons about QCD and its
nonperturbative dynamics beyond calibration work needed for a
better analysis of beauty decays. On the
other it provides a unique portal  to New Physics through up-type quark
dynamics with many, though  not all, experimental features favourable.

In this latter quest only now have we begun to enter promising
territory, namely gaining sensitivity for $x_D$ and $y_D$ values of
order percent and likewise for CP asymmetries.

Without a clearcut theory of New Physics one has to strike a balance
between the requirements of feasibility and the demands of making a
sufficiently large step beyond what is known when suggesting
benchmark numbers for
the experimental sensitivity to aim at. In that spirit I suggest the
following numbers:
(i)
Probe $D^0 - \bar D^0$ oscillations down to
$x_D$, $y_D$ $\sim {\cal O}(10^{-3})$ corresponding to
$r_D \sim {\cal O}(10^{-6} - 10^{-5})$.
(ii)
Search for {\em time dependant} CP asymmetries in
$D^0(t) \to K^+K^-$, $\pi ^+\pi ^-$, $K_S\phi$ down to the
$10^{-4}$ level and in the doubly Cabibbo suppressed mode
$D^0(t) \to K^+ \pi ^-$ to the $10^{-3}$ level.
(iii)
Look for asymmetries in the partial widths for
$D^{\pm} \to K_{S[L]}\pi ^{\pm}$ down to $10^{-3}$ and likewise
in a {\em host} of singly Cabibbo suppressede modes.
(iv)
Analyze Dalitz plots and T-odd correlations etc. with a sensitivity
down to ${\cal O}(10^{-3})$.

Huge amounts of new information on charm dynamics will become available
due to data already taken by FOCUS and SELEX and being taken at the $B$
factories; there
is activity to be hoped for at Compass, BTeV and
LHC-B. And finally there are the `gleam in the eye' activities that
could be performed at a tau-charm factory at Cornell. We can be sure
to learn many relevant lessons from such studies -- and there may be
surprises when we least expect it as expressed by the following
allegory:

`The poor sleeper's impatience'

\noindent A man wakes up at night,

\noindent Sees it is dark outside and falls asleep again.

\noindent A short while later he awakes anew,

\noindent Notices it still to be dark outside and goes
back to sleep.

\noindent This sequence repeats itself a few times

\noindent -- waking up, seeing the dark outside and falling
asleep again --

\noindent Till he cries out in despair:

\noindent "Will there never be daylight?"

\section*{Acknowledgments}
Many thanks go to Tony Sanda and his team for organizing an inspiring
meeting in a beautiful and thus appropriate setting.
This work has been supported by the NSF under the grants
PHY 96-05080 and INT-0089550.

\end{document}